# Kruskal-Wallis Power Studies Utilizing Bernstein Distributions; preliminary empirical studies using simulations/medical studies.

Jeremy S.C. Clark[a*], Piotr Kulig[a], Konrad Podsiadło[a], Kamila Rydzewska[a], Krzysztof Arabski[a], Monika Białecka[b], Krzysztof Safranow[c], Andrzej Ciechanowicz[a]

**Short title:** Kruskal-Wallis Power studies

[a] *Department of Clinical & Molecular Biochemistry, Pomeranian Medical University, Szczecin, Poland.*

[b] *Department of Pharmacokinetics and Therapeutic Drug Monitoring, Pomeranian Medical University, Szczecin, Poland.*

[c] *Department of Biochemistry and Medical Chemistry, Pomeranian Medical University, Szczecin, Poland.*

*To whom correspondence should be addressed:

dr hab. n. med. Jeremy Clark, Department of Clinical & Molecular Biochemistry, Pomeranian Medical University, ul. Powstancow Wlkp. 72, 70-111 Szczecin, Poland.

Tel. 004891 4661490; Fax. 004891 4661492; Email:   jeremyclarkbio@gmail.com,

**Acknowledgments.**

We would like to thank the National Sleep Research Resource for provision of data via data access and user agreement NSRR R24 HL114473: NHLBI. We would like to thank Prof. Torsten Hothorn (University of Zurich, Switzerland) for advice concerning power studies, Tomasz Prystacki and Leszek Domański for the provision of data; and Martin Becker for provision of coding (for empirical Pearson type sampling) for production of simulation data.

**ABSTRACT.**

Bernstein fits implemented into R allow another route for Kruskal-Wallis power-study tool development. Monte-Carlo Kruskal-Wallis power studies were compared with measured power, with Monte-Carlo ANOVA equivalent and with an analytical method, with or without normalization, using four simulated runs each with 60-100 populations (each population with N=30000 from a set of Pearson-type ranges): random selection gave 6300 samples analysed for predictive power. Three medical-study datasets (Dialysis/systolic blood pressure; Diabetes/sleep-hours; Marital-status/high-density-lipoprotein cholesterol) were also analysed. In three from four simulated runs (run_one, run_one_relaxed, and run_three) with Pearson types pooled, Monte-Carlo Kruskal-Wallis gave predicted sample sizes significantly slightly lower than measured but more accurate than with ANOVA methods; the latter gave high sample-size predictions. Populations (run_one_relaxed) with ANOVA assumptions invalid gave Kruskal-Wallis predictions similar to those measured. In two from three medical studies, Kruskal-Wallis predictions (Dialysis: similar predictions; Marital: higher than measured) were more accurate than ANOVA (both higher than measured) but in one (Diabetes) the reverse was found (Kruskal-Wallis: lower; Monte-Carlo ANOVA: similar to measured). These preliminary studies appear to show that Monte-Carlo Kruskal-Wallis power studies based on Bernstein fits might perform better than ANOVA equivalents in many settings (and provide reasonable results when ANOVA cannot be used); and both Monte-Carlo methods appeared considerably more accurate than the analysed analytical version.

# INTRODUCTION.

There are few freely available methods for power analysis using Kruskal-Wallis tests. Some authors have suggested estimation of power and sample size using permutation methods, i.e. generating random datasets of a prespecified distribution with given input parameters. Using permutation, Hecke (2012) compared simulated populations with normal, lognormal and chi-squared distributions and found that Kruskal-Wallis tests compared favorably with analysis of variance (ANOVA) and it has been known for a long time that, if medians are to be tested, the Kruskal-Wallis test is competitive to the *F*-test (Feir-Walsh and Toothaker, 1974).

Many alternative methods have been proposed. Mahoney and Magel (1996) tested four underlying continuous distributions that possessed various location configurations to estimate power for Kruskal-Wallis analyses utilizing bootstrapping techniques to produce power estimates based on empirical cumulative distribution functions. The preferred configuration (the extended average X and Y method) was reliable for three groups (for more groups power was believed to be overestimated).

For continuous data, Fan et al. (2011) did not pursue the average X and Y method, but proposed and tested alternatives methods: the AdjGeneric method and the shift g method - both thought to provide better estimates of power than F tests. The shift g method of Fan et al. (2011) has, as far as we know, not been implemented in R. In 2012, Fan and Zhang (2012) extended their work to encompass ordinal data and concluded that their method performed better than various Monte-Carlo methods if the underlying group distributions were not known (sample sizes needed to be larger than 100).

Several aforementioned algorithms have not so far to our knowledge been implemented in the R statistical platform packages and there appears to be no user-friendly tool freely available for the calculation of power and predicted sample sizes which utilizes the Kruskal-Wallis test. There is also the possibility that Monte-Carlo methods, where predicted distributions are created and then sampled at increasing sample size to determine power, might be more accurate than analytical methods. From an initial pilot study with mean differences between three study groups, a power study can use characteristics of these groups to create simulated populations, randomly sample these and estimate sample size for a statistically significant Kruskal-Wallis test difference. The method proposed and implemented utilizes Bernstein distribution fits to pilot study data. For this we rely heavily on the work carried out by Hothorn including the simulation testing of Bernstein fits (Hothorn et al., 2008, 2017).

For simulated studies, up to 180 simulated populations from different Pearson distributions were analyzed, with different skewness/kurtosis combinations, plus populations with one group with normal moments, plus some analyses where ANOVA was performed on transformed data. For assessments with real data, three reasonably sized medical study datasets were analyzed: blood pressure measurements from a retrospective review of dialysis center medical records; number of hours of sleep/diabetes classes from a multi-center epidemiologic study; and HDL-C measurements/marital status from a multi-center cohort study.

For power comparison the created simulated datasets plus the three medical datasets provided "populations" from which initial samples of small size were drawn for "prediction" of sample size needed, and the populations were sampled at increasing sample size to give "measured" actual sample size at required power. Tools for Monte-Carlo Kruskal-Wallis power studies utilising Bernstein fits are provided for immediate use (Supplemental Files S7, S8).

The main aim of this paper was to perform comparisons between Monte-Carlo ANOVA, an analytical ANOVA and a Monte-Carlo Kruskal-Wallis power study based on Bernstein fits, and between these and measured power, using simulated datasets of various Pearson types, plus three medical study datasets.

**METHODS.**

**Simulation Studies (S1B_Simulation_settings.R).**

Populations for each simulation run were produced with usually 10 000 values for each group and empirically-defined Pearson types (PearsonDS::*rpearson* (Martin and Stefan, 2017)) within tolerance (rather than being drawn from infinite type populations; tolerance coding courtesy of Martin Becker; see Table 1; note that processing was not possible for all types). For each run a particular population variance was applied (Table 1) and mean differences were adjusted to give samples within reasonable time. Instructions for type production were saved in a dataframe "skewdf" and later passed to the list of population dataframes "mydf", in particular note skewdf[ , "ExpectNormal"] had possible values "Norm", "Not-norm" or "KruskalOnly", according to whether it was reasonably possible (within time constraints) to produce samples (or populations) with ANOVA (or ANOVA with transformed variable) requirements, or only with Kruskal-Wallis requirements, respectively. Each population had all groups with the same Pearson type (skewdf1, mydf1), or with one group (G3) with normal moments (skewdf2, mydf2) or population copies of skewdf1/mydf1 with instructions for later normalization (skewdf3, mydf3). skewdf1, skewdf2 and skewdf3 were combined to give skewdf and similarly to give mydf.

Selection criteria for each population were (a) a significant difference (with alphaPOP) by Kruskal-Wallis test between population groups; (b) Kruskal-Wallis power (with alpha) of 100 initialsizeN samples not above "power" (80% or 90%); and (c) for certain population types (ExpectNormal == "Norm" or "Not-norm") ANOVA power (with alpha) of 100 initialsizeN samples not above "power".

**Data - Medical Studies.**

**Study (1). Dialysis study (run_Dial; S1A_Dialysis_settings.R, S9A_Dialysis_data.csv).** Systolic blood pressure measurements (mmHg; N = 15062) with three groups of dialysis centers. Significant differences were previously found among 10 dialysis centers in Poland (Prystacki et al., 2013); here those with highest, medium and lowest blood pressure measurements were deliberately grouped together to give three groups which were statistically significantly different (Supplementary-Files-1,7).

**Study (2). Diabetes study (run_DIAB; S1C_DIABETES_settings.R, S9C_DiabetesData.csv).** Self-reported number of hours of sleep (h/day; N = 13362; parameter "SLPDUR") with three American Diabetes Association diabetes classes (parameter "diabetes2"): 1, no diabetes; 2, pre-diabetes; and 3, treated diabetes; from the Hispanic Community Health Study (https://sleepdata.org/datasets, NHLBI National Sleep Science Resource (Dean et al., 2016); Supplementary-File-2).

**Study (3). Marital study (run_Mar; S1D_Marital_settings.R, S9D_MaritalData.csv).** Serum high-density-lipoprotein-cholesterol levels (HDL-C; mg/dL; N = 5079, parameter "hdl") and marital status (parameter "mstat": 1, married; 2, widowed; 3, divorced/separated

(the small categories 4, never married and 8, unknown/refused, were removed); from the Sleep Heart Health Study (https://sleepdata.org/datasets, NHLBI National Sleep Science Resource (Quan et al., 1997); Supplementary-File-3).

**Coding organisation.**

The S2A_SAMPLE_ MASTER.R file was run in R (R 4.0.3 GUI 1.73) to stitch a settings file (see Studies above) with the S2B_Sample_Coding_ section_2.R file to produce a SampleCoding file which was run on a single processor (e.g. Macbook Air, macOS Catalina 10.15.7, processor 1.8 GHz, memory 5 GB).

In this article "TRIAL" is used to refer to power studies from samples at initialsizeN whereas "MEASURE" is used to refer to measurements of power at increasing sample sizes. To produce many SCRIPT files (each analysing a number of samples) the S2C1_POWER_ MASTER.R file stitched the S3_Load_Coding.R file to a power coding file (S4_Power_Coding_ parts1and2_Krus.R, S4_Power_ Coding_part3_ANOVA.R or S4_Power_Coding_ part4_MEASURE.R). SCRIPT files were either for: part1: Kruskal Wallis TRIAL division method (KRTDIV); part2: Kruskal Wallis TRIAL duplicate method (KRTDUP); part3: ANOVA TRIAL or part4: Kruskal-Wallis MEASURE.

An Eagle Cluster (CPU E5-2697 v3 @ 2.60GHz haswell; Ethernet 100/1000; InfiniBand FDR; Lustre storage; Operating System: CentOS Linux release 7.6.1810) at the Poznan Supercomputing and Networking Center, eagle.man.poznan.pl, Poznan, Poland was accessed using R parallel::*makeCluster* (R Core Team, 2020) and doParallel :: *registerDoParallel* (Microsoft and Weston, 2019) and the SCRIPT file main loop was implemented as a foreach::*foreach* (Microsoft and Weston, 2020) function with set seeds distributed using [doRNG] (Renaud, 2020) functions *registerDoRNG*, *.options.RNG* and *%dorng%*.

Samples files, script files, data (for medical studies), and slurm files (e.g. S5_aaSLURM _sim_run_two_part2.sl) were passed (using Forklift 3.4.4) to the cluster and run for 4 hours maximum. Occasionally (through timeout or memory failure) it was necessary to run the S2C2_POWER_MASTER_SUBSET.R file to produce SCRIPT files for smaller numbers of samples. Result ("datastore") files were collated using the S6_Collation_Coding.R file. Graphics were produced using ggplot2 (Wickham, 2016) (S10_GRAPH_NON _PARAMETRIC.R). All references for R libraries are given at end of S1A.

**Sample-Selection Coding, S2B_Sample_Coding_ section_2.R.**

For all studies from each population 10 or 30 samples were created without significant group differences by Kruskal-Wallis test (plus by ANOVA unless relaxed or population skewdf[ , "ExpectNormal"] == "KruskalOnly"). Sample sizes (termed initialsizeN, all groups together) were chosen (arbitrarily) at around 1/3 sample size of that for preliminary measured power of 90%. With skewdf "Norm" samples were normally distributed (R [stats] *shapiro.test*); with skewdf "Not-norm" transformed samples (usually with jtrans::*jtrans* or lamW::*lambertW0*) were normally distributed; with skewdf "KruskalOnly" no requirement for normality was applied. Settings for all runs are found in Table 1.

**Predicted Sample Sizes (TRIALs).**

For Kruskal-Wallis predicted power, Bernstein distributions were fitted to samples using R [mlt] (Hothorn et al., 2017) functions: *numeric_var*, *as.basis*, *ctm* and *mlt* (alternative methods internal to the fitting: Duplication (KRTDUP) method: data duplicated or Division (KRTDIV) method: data ranfomly split into two). The distributions were sampled (using

*simulate*) at increasing sample sizes (starting from initialsizeN+1) until required power. For ANOVA (with White adjustment) predicted power, normal distributions were fitted and sampled. Effect sizes used Vargha and Delaney's A statistic.

**Actual Power (MEASURE).**

Actual sample sizes needed to obtain a particular power were estimated by analysing random samples at increasing sample size (starting with samplesizeN+1). As with the Trials the samples drawn had the same (rounded) proportions as in the population.

**Analytical.** Analytical ANOVA power used stats::*power.anova.test*; coding in S4_Power_Coding_ part4_ MEASURE.R.

**Statistics.** All tests were two-tailed. Alpha for the Medical studies set at 0.05. Differences between ANOVA and Kruskal-Wallis "trial" or "measured" power were determined using Brown-Mood median tests. Non-parametric aligned rank transform was also used to compare parameters (TRIALs, MEASURE, Analytical) and results are given in S10_GRAPH_NON_PARAMETRIC.R or S12_Medians_Moods.R. (Note that parametric comparisons were also attempted: Poisson, negative binomial and zero-inflated models, but these failed mostly due to excessive overdispersion.)

**Code availability.** All code is provided.

**Data availability.** Simulation studies: sample (mydfsample) values are stored in datastore files, which also give many other statistics, in Supplemental files S3B....datastore_noPOP.RData. Dialysis Study (1): S9A_Dialysis_data.csv, numeric population values also given in S3A_Dialysisrun_Dial_datastore.RData; Diabetes Study (2): the Hispanic Community Health Study from https://sleepdata.org/datasets, NHLBI National Sleep Science Resource (Dean et al., 2016); parameters "diabetes2" and "SLPDUR". Marital

Study (3): the Sleep Heart Health Study; as for (2), parameters "hdl" and "mstat". All coding and Supplemental files are found at https://github.com/ Abiologist/Power.git (or via arxiv: 110.11676v1).

**RESULTS.**

The main results from simulations studies are shown in Table 2 and Figs 1:4 which can be generated by running S10_GRAPH_NON_PARAMETRIC.R. The two internal methods (duplication, KRTDUP or division, KRTDIV) for Monte-Carlo Kruskal-Wallis prediction gave similar results for all analyses and only KRTDIV is shown in these Figures. The three prediction methods KRTDIV, Monte-Carlo ANOVA (ANT) and analytical ANOVA power (Analytical) were compared with measured power from ANOVA (ANM) and/or Kruskal-Wallis (KRM). The best methods were presumed to be those with medians most similar to KRM and ANM (ANM gave very similar distributions to KRM but failed more often; note that modes shown in Supplemental Figures can be deceptively different from medians or means).

From simulated run_three (Figs 1, OR1) with variance 20, group means ~98 to ~100, Pearson types 0 and VII (Table 2), KRTDIV pooled results (from all Pearson distributions) gave a median predicted sample size (all values in this section: median $\pm$ median absolute deviation) of $221 \pm 70$ which was 8% lower that for measured power (KRM or ANM: $241 \pm 30$). This indicates that the Monte-Carlo Kruskal-Wallis power studies were fairly accurate but non-conservative. Monte-Carlo ANOVA (ANT) gave a pooled median of $281 \pm 110$ i.e. 17% higher and was therefore less accurate overall than the Kruskal-Wallis method and conservative (note the lower median absolute deviation for KRTDIV than for ANOVA).

The analytical method (see Fig OR1, Table 2) gave higher values (357 ± 150). For these settings the numbers of fails were negligible (0.4%).

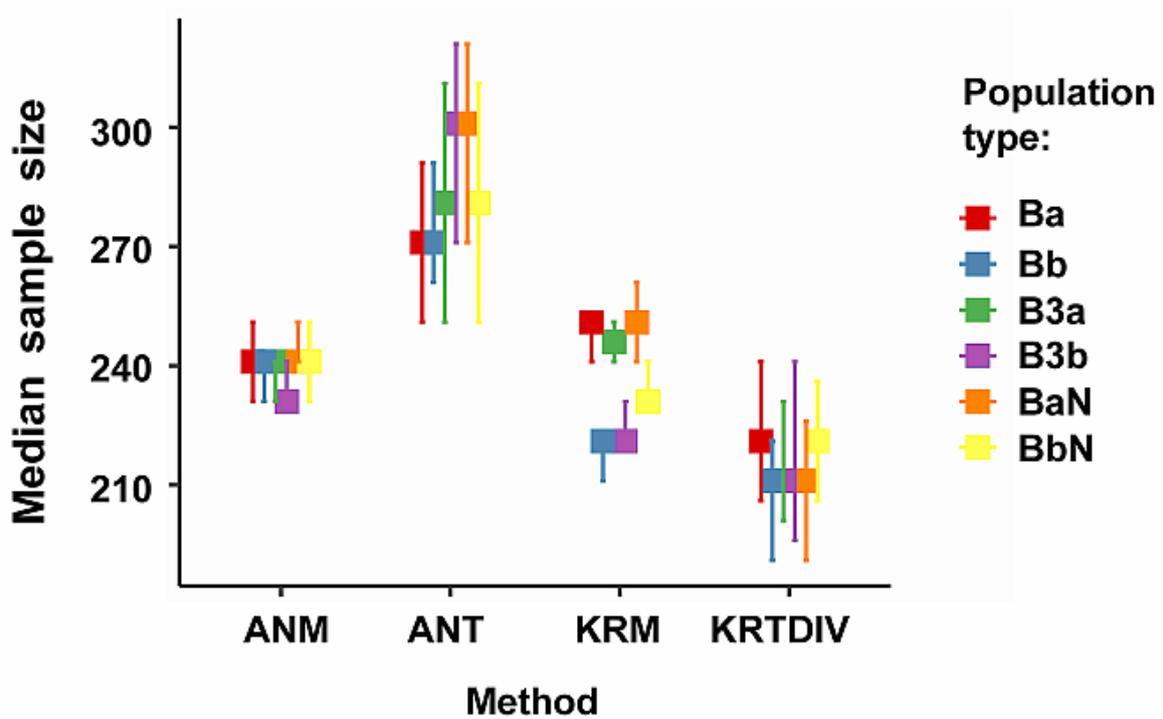

**Figure 1. Simulation run_three**: group means: 100, 100-1, 100-2; variance 20. Medians of measured (ANM: ANOVA, KRM: Kruskal-Wallis) and Monte-Carlo predicted (ANT: ANOVA, KRTDIV: Kruskal-Wallis) sample sizes for 80% power. Error bars: 95% confidence intervals. For run settings, Methods and Population type definitions, see Tables 1,2.

From simulated run_one (Figs 2, OR5) with variance 5, group means ~99 to ~100 with many Pearson types (Table 2), KRTDIV pooled results (from all Pearson distributions) gave a median predicted sample size of 211 $\pm$ 80, 9% lower that for measured power (KRM or ANM: 231$\pm$ 30). Again the Monte-Carlo Kruskal-Wallis method was fairly accurate but non-conservative. Monte-Carlo ANOVA (ANT) gave a pooled median of 281 $\pm$ 100 i.e. 22% higher and was therefore less accurate overall than the Kruskal-Wallis method and conservative. The analytical method (see Fig OR2) gave higher values (360 $\pm$ 153). For these settings the numbers of fails were not negligible (34%) due to ANOVA failures. (Note that the median absolute deviation for KRTDIV was less than for the ANOVA methods.)

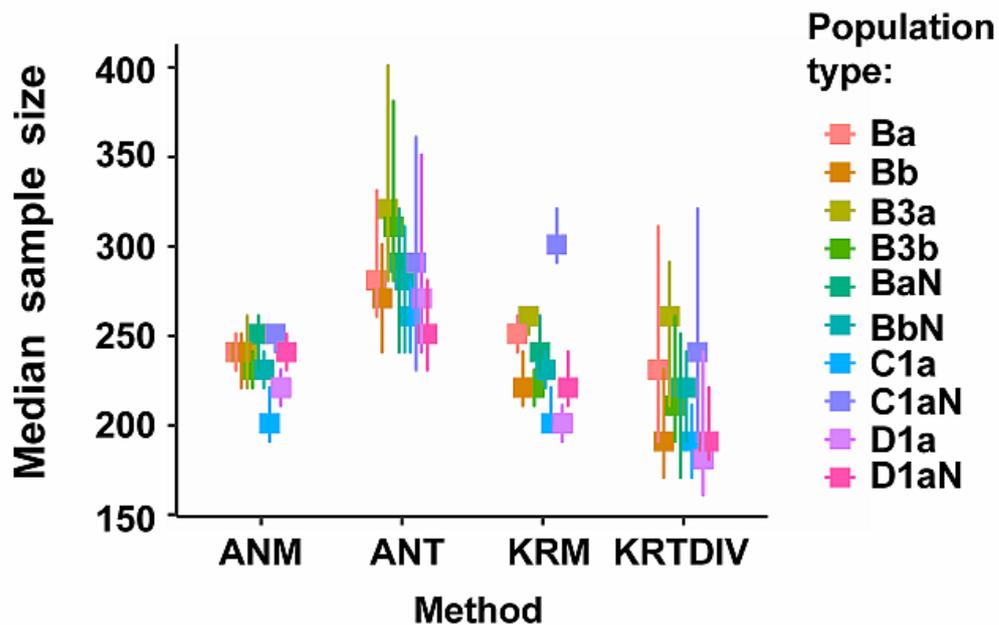

**Figure 2. Simulation run_one**: group means: 100, 100-0.5, 100-1; variance 5. Medians of measured (ANM: ANOVA, KRM: Kruskal-Wallis) and Monte-Carlo predicted (ANT: ANOVA, KRTDIV: Kruskal-Wallis) sample sizes for 80% power. Error bars: 95% confidence intervals. For run settings, Methods and Population type definitions, see Tables 1,2.

Simulated run_onerelaxed i.e. similar to run_one but with samples chosen relaxed from ANOVA assumptions (Figs 3, OR6) with the same range of means and variance as run_one and many Pearson types (Table 2), might be regarded as giving the most realistic simulations. From run_onerelaxed KRTDIV pooled results (from all Pearson distributions) gave a median predicted sample size of $201 \pm 70$, 11% lower that for average measured power ($226 \pm 30$). Again the Monte-Carlo Kruskal-Wallis method was fairly accurate but non-conservative. Monte-Carlo ANOVA (ANT) gave a pooled median of $271 \pm 105$ i.e. 20% higher and was therefore less accurate overall than the Kruskal-Wallis method and conservative (note it could be argued it shouldn't be used at all for many of these populations !). The analytical method (see Fig OR6) gave higher values ($344 \pm 155$). For these settings the numbers of fails were negligible (0.1%). (Note that the median absolute deviation for KRTDIV was less than for the ANOVA methods.)

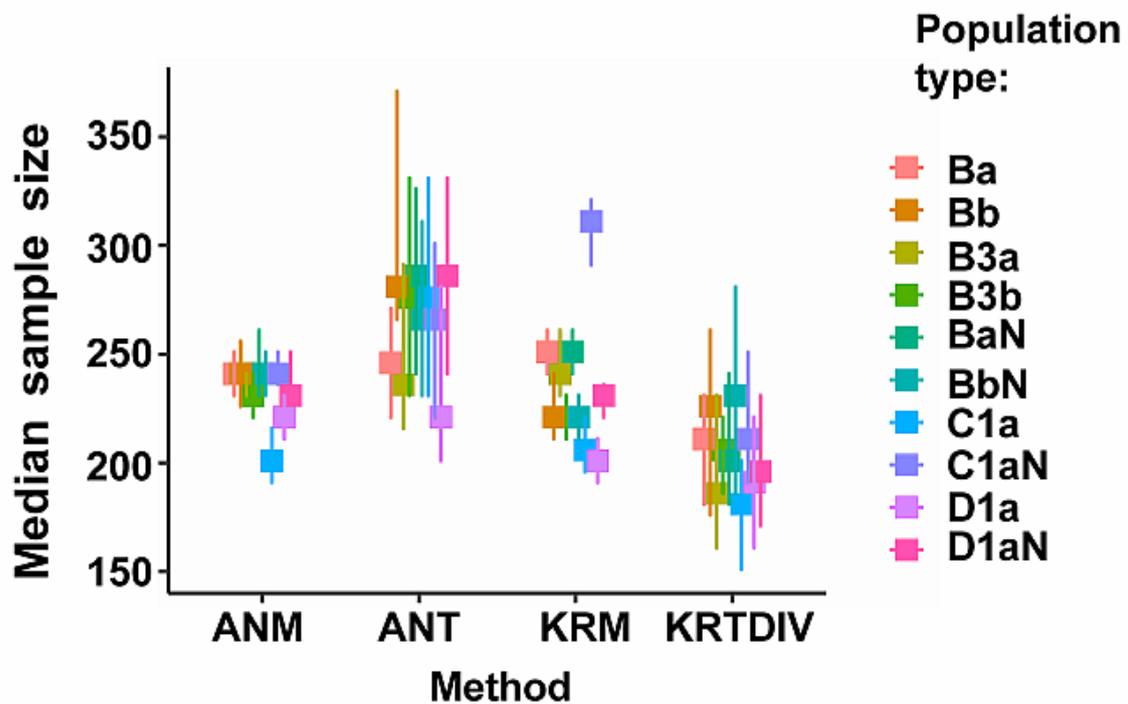

**Figure 3. Simulation run_one relaxed**: group means: 100, 100-0.5, 100-1; variance 5; relaxed = samples chosen regardless of whether they satisfied ANOVA assumptions. Medians of measured (ANM: ANOVA, KRM: Kruskal-Wallis) and Monte-Carlo predicted (ANT: ANOVA, KRTDIV: Kruskal-Wallis) sample sizes for 80% power. Error bars: 95% confidence intervals. For run settings, Methods and Population type definitions, see Tables 1,2.

From simulated run_two (Figs 4, OR7) with variance 5, group means ~99.45 to ~100 with many Pearson types (Table 2), KRTDIV pooled results (from all Pearson distributions) gave a median predicted sample size of 456 ± 180, 29% lower that for average measured power (646

± 90). This time the Monte-Carlo Kruskal-Wallis method was less accurate and non-conservative. Monte-Carlo ANOVA (ANT) gave a pooled median of 581 ± 240 i.e. 9% higher and was therefore more accurate than the Kruskal-Wallis method and conservative. The analytical method (see Fig OR7) gave higher values (827 ± 389). For these settings the numbers of fails were not negligible (34%) due to ANOVA failures. (Note that the median absolute deviation for KRTDIV was less than for the ANOVA methods.)

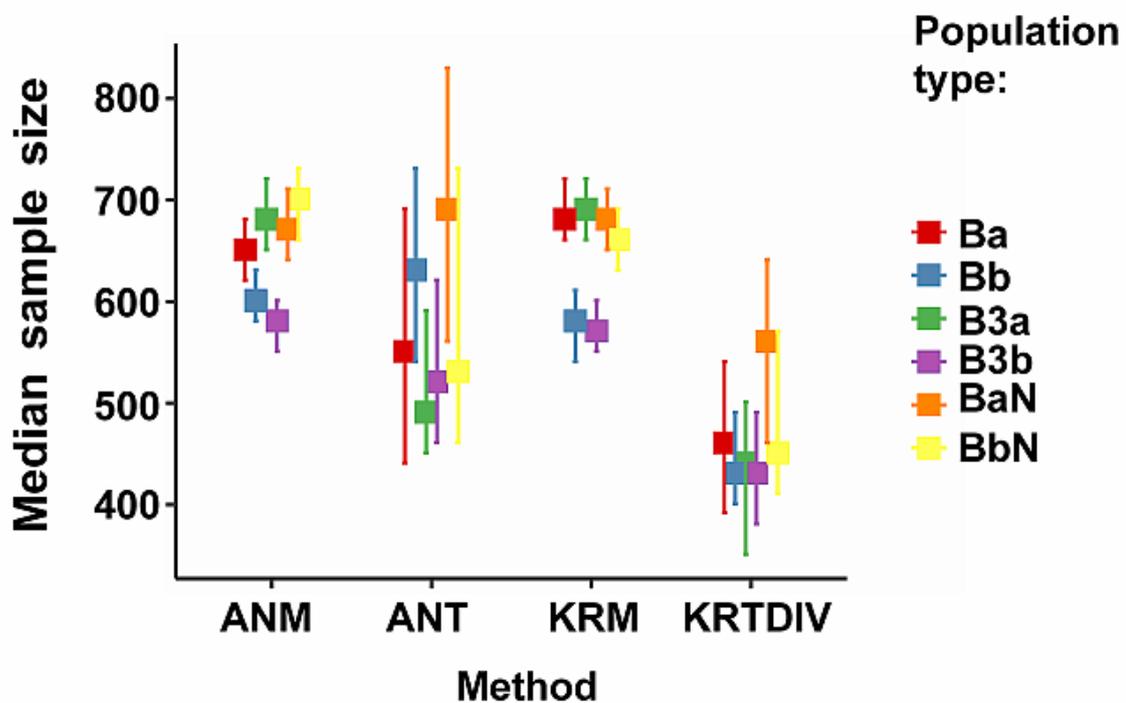

**Figure 4. Simulation run_two**: group means: 100, 100-0.3, 100-0.55; variance 5. Medians of measured (ANM: ANOVA, KRM: Kruskal-Wallis) and Monte-Carlo predicted (ANT: ANOVA, KRTDIV: Kruskal-Wallis) sample sizes for 80% power. Error bars: 95% confidence intervals. For run settings, Methods and Population type definitions, see Tables

1,2.

For those settings including Pearson types which were difficult or impossible to process using ANOVA (Table 3), the Kruskal-Wallis method appeared to do rather well for run_one but not so for run_two, when compared with measured power (KRM).

The diabetes study had data with means ~7.8 to ~8, variance 2, Pearson type 0 and no power study method appeared to perform well (Fig OR2, Table 4); note that ANM failed completely). The median predicted power for KRTDIV (2300 $\pm$ 600) was significantly (26%) lower than from KRM (3100 $\pm$ 300) whereas the median predicted power from ANT (3300 $\pm$ 1000) was not significantly different than KRM (see Table 4). Analytical ANOVA gave a wide range of predictions (5100 $\pm$ 3200).

The dialysis study had data with means ~127 to ~135, variance 427, Pearson type 0, and power studies appeared to perform well (Fig OR3; but note ANM failed completely; Table 4). The median for KRTDIV (570 $\pm$ 200) was identical to that for KRM (570 $\pm$ 100), whereas the median from ANT (770 $\pm$ 300) was significantly (35%) higher (Table 4), again showing that the Kruskal-Wallis method was more accurate. The analytical method gave a high value (670 $\pm$ 360).

With the Marital study (Fig OR4) sample production had to use a relaxed method (not analysing ANOVA), and all samples were analysed with ANOVA (Johnson transformed) and Kruskal-Wallis methods. All prediction methods were significantly conservative i.e. with predicted sample sizes significantly higher than ANM (460 $\pm$ 70) or KRM (390 $\pm$ 40).

KRTDIV (760 $\pm$ 160) performed slightly better than ANT (950 $\pm$ 260) and Analytical (2100 $\pm$ 970) performed considerably worse (Fig OR4; Table 4).

Result ("datastore") files contain considerable statistics (full parameter list in S11_Parameter_ Definitions.xlsx) including empirical moments, Vargha and Delaney's A effect sizes, and all p values and rank orders for final tests. Minimum detected effect sizes (predicted and measured) were similar for ANOVA and Kruskal-Wallis power studies.

**DISCUSSION.**

Efficient study replication should involve power studies to estimate minimal sample sizes while minimising false negative frequencies.

The Kruskal-Wallis test is commonly applied to non-parametric studies with continuous variables and a derived power study theoretically has broad application. Bernstein fits have recently been implemented in R and therefore provide a basis by which Monte-Carlo Kruskal-Wallis prediction can be formulated and, as well as having broader applicability, appear overall to be more accurate than ANOVA (both Monte-Carlo or analytical) methods even when conditions for ANOVA are fulfilled.

With the datasets analysed in the present study using simulated or medical data, Monte-Carlo Kruskal-Wallis power prediction did not seem to be at any particular disadvantage when compared with Monte-Carlo ANOVA (and both Monte-Carlo methods appeared considerably more accurate than the Analytical method analysed). Monte-Carlo Kruskal-Wallis under most circumstances was more accurate, even when normal distribution was demanded (run_one and run_three) and detected similar minimum effect sizes. (Note that ties in the data are allowed.) (It is possible, as suggested by run_two and the diabetes study, that in cases where the group means are very similar the Kruskal-Wallis method might be less accurate than Monte-Carlo ANOVA (although further studies are needed to show that this is indeed the case).

Overall, from these preliminary studies, it appears that the Monte-Carlo Kruskal-Wallis power studies, based on Berstein fits, might be able to be used in preference to ANOVA, and could probably be used if ANOVA failed (see Table 3), or if assumptions were in doubt, over a much wider range of different samples. Additionally, both Monte-Carlo methods were more accurate overall than the analytical version compared.

Further large-scale testing is needed to show the scope of conditions under which the accuracy of the Kruskal-Wallis method falls, and also for example with truncated data, bimodal distributions or with further examples of unequal group sizes.

As in Fan et al. (2011), in the present study for the simulations the group sizes were equal, and it might well be the case that the Kruskal-Wallis method might perform better than ANOVA overall if sizes of individual groups were allowed to differ. It is unknown whether, as in Fan and Zhang (2012), the method will perform poorly with small group sizes.

The fact that the Kruskal-Wallis method compared favorably with ANOVA both with samples from real datasets and from simulated populations with moderate skewness leads us to suggest that, in the absence of a freely available alternative, the method can be used immediately, possibly in preference to ANOVA methods.

It will be intriguing to see the extent to which the Kruskal-Wallis method has advantage over ANOVA under a larger range of conditions and how the method compares with the (rather numerous) possible power-study alternatives (see Introduction and (Fan and

Zhang, 2012; Hecke, 2012; Mahoney and Magel, 1996)). Further large-scale testing against these and other possible methods is advised, together with a larger range of simulated data.

We present these results and implemented code to provide simulation and empirical evidence for utility of a freely available Monte-Carlo Kruskal-Wallis power method which utilizes Bernstein distribution fits as developed by Hothorn et al. (2008, 2017). While the method can be used immediately, further testing is advised to determine the scope under which this method has advantage over ANOVA or where accuracy falls.

**Limitations:**

1) Analysis of further empirical datasets is needed in order to confirm the finding that the Monte-Carlo Kruskal method has some advantages over other methods.

2) Although the Kruskal-Wallis method can be used immediately, further large-scale simulation testing would allow delimitation of the conditions under which accuracy falls.

**Conclusions:**

This preliminary study has shown that, with samples for which ANOVA can also be used, the Monte-Carlo Kruskal-Wallis method was overall more accurate than Monte-Carlo ANOVA. Additionally, the Monte-Carlo Kruskal-Wallis power study method based on Bernstein distributions theoretically has much broader application than ANOVA and a small amount of empirical evidence was provided (in comparison to "measured" power) that this is indeed the case. Of some concern is that both Monte-Carlo methods were found to be considerably more accurate than the analytical method used, which also needs to be confirmed with a larger study. Further large-scale testing is needed in order to delimit the conditions under which accuracy falls.

**Supplementary Materials.**

The reader is referred to the on-line Supplementary Materials for complete R coding plus example SLURM coding, datastores for later studies containing many statistics from method comparisons, and data for the Dialysis study. **Supplemental Files: S1A, S1B, S1C, S1D:** settings files for medical and simulation studies. **S2A, S2B:** sample master and coding. **S2C:** power and power subset masters. **S3:** load coding. **S3A, S3B, S3C, S3D:** datastores. **S4:** power coding files. **S5:** example slurm coding. **S6:** collation coding. **S7, S8:** Monte-Carlo tools - provided as is ! **S9A:** Dialysis data. **S10:** graphical coding. **S11:** parameter definitions for datastores; **S12:** summary medians and Brown-Moods tests. **Supplemental Figures:** OR1:OR7. All supplemental files, figures and all coding is found at https://github.com/Abiologist/Power.git (or via arxiv: 110.11676v1).

**Statements and Declarations.**

**Authors' contributions:** JSCC contributed to concept/design, data analysis/interpretation, drafting the article; AC, MB to data collection, critical revision of article, PK, KP, KR to drafting of article, statistical analysis; KP, KR, KA, KS to non-parametric statistical analysis, power studies.

**Funding:** This study was funded entirely by the Pomeranian Medical University, Szczecin, Poland.

**Competing Interests.** The Authors declare that there is no conflict of interest

**Table 1. A.** Population definitions (POPtypes) with Pearson types (see "POPtype" in skewdf or mydfsample in Samples files). Populations with POPtype suffix including "3", e.g. B3a, had instructions for normalisation. **B.** Run absolute effect sizes and variances for all data and groups G1, G2 and G3. (run_onerelaxed created similarly to run_one but without reference to ANOVA assumptions.) Populations with POPtype "N" suffix had group G3 with normal moments.

| A. POPtype | Skewness | Kurtosis | Pearson type | runs including this POPtype |
|---|---|---|---|---|
| Aa | 0 | 1.1 | II | not analysed |
| Ba or B3a | 0 | 3 | 0 | all sim. runs, run_Dial, run_DIAB |
| Bb or B3b | 0 | 5 | VII | all sim. runs |
| C1a or C3a | 1 | 5 | V | run_one, run_two, run_Mar |
| C1b or C3b | 1 | 3 | I | not analysed |
| D1a or D3a | 0 | 10 | VII | run_one, |
| D1b or D3b | 1 | 10 | IV | run_one, run_two |
| E1a or E3a | 2.14 | 10 | III | run_one |
| E1b or E3b | 2.14 | 15 | IV | not analysed |
| F | 3 | 15 | I | not analysed |

sim. = simulation

| B. Run | Population means** (medians): G1; G2; G3; Total | Variance (MAD) G1; G2; G3; Total | POPtypes*** |
|---|---|---|---|
| run_one | 100; 100-0.5; 100-1 | 5, 5, 5, n.d. | Ba Bb B3a B3b BaN BbN C1a C1aN C3a D1a D1aN D1b D1bN D3a D3b E1a E1aN E3a |
| run_one - relaxed | 100; 100-0.5; 100-1 | 5, 5, 5, n.d. | Ba Bb B3a B3b BaN BbN C1a C1aN C3a D1a D1aN D1b D1bN D3a D3b |
| run_two | 100; 100-0.3; 100-0.55 | 5, 5, 5, n.d. | Ba Bb B3a B3b BaN BbN C1a C1aN C3a D1b D1bN D3b |
| run_three | 100; 100-1; 100-2 | 20, 20, 20, n.d. | Ba Bb B3a B3b BaN BbN |
| run_Dial | 135 (135); 132 (130); 127 (130); 132 (130) | 452 (15); 341 (10); 434 (10); 427 (10) | Ba |
| run_DIAB | 8.02 (8); 7.82 (7.86); 7.92 (8); 7.92 (8) | 1.99 (0.929); 1.96(0.857); 2.31 (1); 2.05 (1) | Ba |
| run_Mar | 50.0 (47); 55.4 (53); 52.3 (50); 50.7 (48) | 247 (10); 231 (10); 249 (10); 248 (10) | C1a |

MAD = median absolute deviation (R *mad(, constant=1)*); n.d.=not determined; **simulation settings (medians and totals not determined) or measured from medical data. *** POPtypes set for simulations or approximated for medical data.

**Table 2.** Predicted and measured sample sizes using Monte-Carlo or analytical ANOVA, or Kruskal-Wallis tests, using samples derived from simulated populations. Results from all Pearson types pooled.

| Simulation run (see Table 1), n | Type. | Estimates for 80% power (Median ± MAD)* | | | |
|---|---|---|---|---|---|
| | | Monte-Carlo ANOVA: ANT, predicted; ANM, measured | Monte-Carlo Kruskal-Wallis: KRTDIV, predicted; KRM, measured | Monte-Carlo Kruskal-Wallis: KRTDUP, predicted | Analytical ANOVA |
| run_one, N = 1800 | Predicted | 280[abc] ± 100 n=1193 | 210[bde] ± 80 n=1193 | 210 ± 70 n=1193 | 360 ± 150 n=1188 |
| | | 280 ± 100 n=1395 | 200 ± 70 n=1798 | 200 ± 70 n=1797 | 360 ± 140 n=1390 |
| | Measured | 230[ad] ± 30 n=1193 | 230[ce] ± 30 n=1193 | | |
| | | 230 ± 30 n=1200 | 220 ± 40 n=1800 | | |
| run_one relaxed, N = 1500 | Predicted | 270[fgh] ± 100 n=1493 | 200[gij] ± 70 n=1493 | 200 ± 70 n=1493 | 340 ± 150 n=1486 |
| | | 270 ± 100 n=1495 | 200 ± 70 n=1498 | 200 ± 70 n=1498 | 340 + 150 n=1490 |
| | Measured | 230[fi] ± 30 n=1493 | 220[hj] ± 30 n=1493 | | |
|  |  | 230 ± 30 n=1500 | 220 ± 30 n=1500 |  |  |
| run_two, N = 1200 | Predicted | 580$^{kmn}$ ± 240 n=788 | 460$^{mpr}$ ± 180 n=788 | 450 ± 180 n=788 | 820 ± 390 n=782 |
|  |  | 580 ± 240 n=893 | 450 ± 190 n=1185 | 450 ± 180 n=1184 | 820 ± 390 n=887 |
|  | Measured | 640$^{kp}$ ± 90 n=788 | 650$^{nr}$ ± 90 n=788 |  |  |
|  |  | 640 ± 90 n=800 | 640 ± 100 n=1200 |  |  |
| run_three, N = 1800 | Predicted | 280$^{stu}$ ± 100 n=1790 | 220$^{tvw}$ ± 70 n=1790 | 220 ± 80 n=1789 | 350 ± 150 n=1783 |
|  |  | 280 ± 110 n=1794 | 210 ± 70 n=1794 | 220 + 80 n=1794 | 350 + 150 n=1785 |
|  | Measured | 240$^{sv}$ ± 30 n=1790 | 240$^{uw}$ ± 30 n=1790 |  |  |
|  |  | 240 ± 30 n=1800 | 240 ± 30 n=1800 |  |  |

N = original number of samples; *Methods: ANT, ANM, KRTDIV, KRTDUP and KRM: definitions in text. Medians and MAD = median absolute deviation around the median (R *mad(constant = 1)*) to two significant figures: first numbers are for statistical comparisons between ANT, ANM, KRTDIV or KRM (all values removed when NA found in either of these), second numbers are without reference to NAs in other parameters. Bonferroni-adjusted comparisons with same letter significantly different at p<0.05, other comparisons not significantly different: $^{abfgprstuv}$p<1.32x10$^{-15}$; $^{a}$Z=8.23; $^{b}$Z=8.88; $^{c}$Z=7.78, p=4.40x10$^{-14}$; $^{d}$Z=4.18, p=0.000148; $^{e}$Z=-3.4387, p=0.00301; $^{f}$Z=8.56; $^{g}$Z=9.74; $^{h}$Z=6.65,

p=1.72x10$^{-10}$; $^i$Z=7.77, p=4.80x10$^{-14}$; $^j$Z=-6.04, p=9.27x10$^{-09}$; $^k$Z=-3.63, p=0.00173; $^m$Z=5.44, p=3.21x10$^{-07}$; $^n$Z=-4.03, p=0.000335; $^p$Z=12.0; $^r$Z=-11.9; $^s$Z=9.86; $^t$Z=10.8; $^u$Z=10.7; $^v$ Z=9.08; $^w$Z=-7.56, p=2.36x10$^{-13}$.

**Table 3.** Predicted and measured sample sizes using Monte-Carlo Kruskal-Wallis tests, using samples derived from simulated populations for which ANOVA would be either difficult or impossible to process. Results from all Pearson types pooled from data for which ANT sample sizes were NA.

| Simulation run, N | Type. | Estimates for 80% power (Median ± MAD) | |
|---|---|---|---|
| | | Monte-Carlo Kruskal-Wallis: KRTDIV or KRM | Monte-Carlo Kruskal-Wallis: KRTDUP |
| run_one, N = 1800 | Predicted | 160 ± 50 n=405 | 160 ± 50 n=404 |
| | | 160 ± 50 n=405 | 160 ± 50 n=404 |
| | Measured | 150 ± 40 n=405 | |
| | | 150 ± 40 n=405 | |
| run_two, N = 1200 | Predicted | 440[a] ± 190 n=299 | 440 ± 190 n=297 |
| | | 440 ± 190 n=299 | 450 ± 200 n=299 |
| | Measured | 670[a] ± 130 n=299 | |
| | | 670 ± 130 n=307 | |

N = original number of samples; Medians and MAD = median absolute deviation around the median (R *mad(constant = 1)*) to two significant figures: first numbers are for statistical comparisons between KRTDIV and KRM (all values removed when NA found in KRTDIV or KRM), second numbers are without reference to NAs in KRTDIV, KRTDUP or KRM. run_one: no significant difference between KRTDIV and KRM medians by Moods tests;

run_two [a]Z=-7.36, p=1.89x10$^{-13}$.

**Table 4.** Predicted and measured Monte-Carlo sample sizes using ANOVA or Kruskal-Wallis tests, using samples derived from Medical study result datasets.

| Study and source. | Type. | Estimates for 90% power (Median ± MAD) | | | | | |
|---|---|---|---|---|---|---|---|
| | | ANOVA, | | | Kruskal-Wallis***, | | |
| | | Sample sizes | Min. effect size** | No. fails | Sample sizes | Min. effect size** | No. fails |
| (1) Systolic blood pressure; 3 groups of dialysis centers. $N_D$ = 15062; (Prystacki et al., 2013). n=197/200. | Predicted, | 770 ± 300[ab] | 0.56 ± 0.021 | 3 | 570 ± 200[a] | 0.56 ± 0.024 | 1 |
| | Measured, | n.c. | n.c. | n.c. | 570 ± 100[b] | 0.58 ± 0.011 | 0 |
| (2) Number of hours of sleep; 3 diabetes classes. $N_I$ = 13362; (Dean et al., 2016). n=79 / 100. | Predicted, | 3300 ± 1000[c] | 0.52 ± 0.014 | 17 | 2300 ± 600[cd] | 0.52 ± 0.013 | 13 |
| | Measured, | n.c. | n.c. | n.c. | 3100 ± 300[d] | 0.53 ± 0.0031 | 0 |
| (3) High-density-lipoprotein cholesterol; 3 | Predicted, | 950 ± 260[efg] | 0.50 ± 0.028 | | 760 ± 160[fhj] | 0.50 ± 0.031 | |

| marital status classes*. $N_I = 5079$; (Quan et al., 1997) n=185/200; | Measured, | $460 \pm 70^{ehi}$ | $0.52 \pm 0.022$ | $390 \pm 40^{gij}$ | $0.52 \pm 0.022$ |

ANOVA = analysis of variance; MAD = median absolute deviation; Min. = minimum; No. fails = number of failures; Predicted: sample sizes with 90% power from predicted distributions created from initial samples with (arbitrarily) approximately one third sample size. Measured: sample sizes with power measured at 90% from data randomly selected from datasets at increasing sample size. n = number of times prediction or measurement carried out. $N_D$ = number of dialysis treatments; $N_I$ = number of individuals. All results to two significant figures. Mood's test (R [coin] *median_test*) global values (2 df): (1) $chi^2 = 29.4$, $p=4.07 \times 10^{-7}$ (2) $chi^2=9.13$, $p=0.0104$ (3) $chi^2=506$, $p<2.2 \times 10^{-16}$; Bonferroni-adjusted comparisons with same letter significantly different at p<0.05, other comparisons not significantly different: [a]$Z=3.46$; $p=0.00171$; [b]$Z=4.65$, $p=1.00 \times 10^{-5}$; [c]$Z=3.65$, $p=0.000263$; [d]$Z=3.49$, $p=0.001451$; [e]$Z=13.6$, $p<1.32 \times 10^{-15}$; [f]$Z=4.26$, $p=0.000124$; [g]$Z=18.2$, $p<1.32 \times 10^{-15}$; [h]$Z=-12.8$, $p=<1.32 \times 10^{-15}$; [i]$Z=7.38$, $p=9.55 \times 10^{-13}$; [j]$Z=18.0$, $p<1.32 \times 10^{-15}$. *after Johnson transformation. **These are medians of minimum effect sizes detected or measured for each trial. ***Division method i.e. KRTDIV. n.c. = not calculable e.g. sample size reached maximum.

**FIGURE LEGENDS.**

**Figure 1. Simulation run_three**: group means: 100, 100-1, 100-2; variance 20. Medians of measured (ANM: ANOVA, KRM: Kruskal-Wallis) and Monte-Carlo predicted (ANT: ANOVA, KRTDIV: Kruskal-Wallis) sample sizes for 80% power. Error bars: 95% confidence intervals. For run settings, Methods and Population type definitions, see Tables 1,2.

**Figure 2. Simulation run_one**: group means: 100, 100-0.5, 100-1; variance 5. Medians of measured (ANM: ANOVA, KRM: Kruskal-Wallis) and Monte-Carlo predicted (ANT: ANOVA, KRTDIV: Kruskal-Wallis) sample sizes for 80% power. Error bars: 95% confidence intervals. For run settings, Methods and Population type definitions, see Tables 1,2.

**Figure 3. Simulation run_one relaxed**: group means: 100, 100-0.5, 100-1; variance 5; relaxed = samples chosen regardless of whether they satisfied ANOVA assumptions. Medians of measured (ANM: ANOVA, KRM: Kruskal-Wallis) and Monte-Carlo predicted (ANT: ANOVA, KRTDIV: Kruskal-Wallis) sample sizes for 80% power. Error bars: 95% confidence intervals. For run settings, Methods and Population type definitions, see Tables 1,2.

**Figure 4. Simulation run_two**: group means: 100, 100-0.3, 100-0.55; variance 5. Medians of measured (ANM: ANOVA, KRM: Kruskal-Wallis) and Monte-Carlo predicted (ANT: ANOVA, KRTDIV: Kruskal-Wallis) sample sizes for 80% power. Error bars: 95% confidence intervals. For run settings, Methods and Population type definitions, see Tables

1,2.

**Supplemental Figures.**

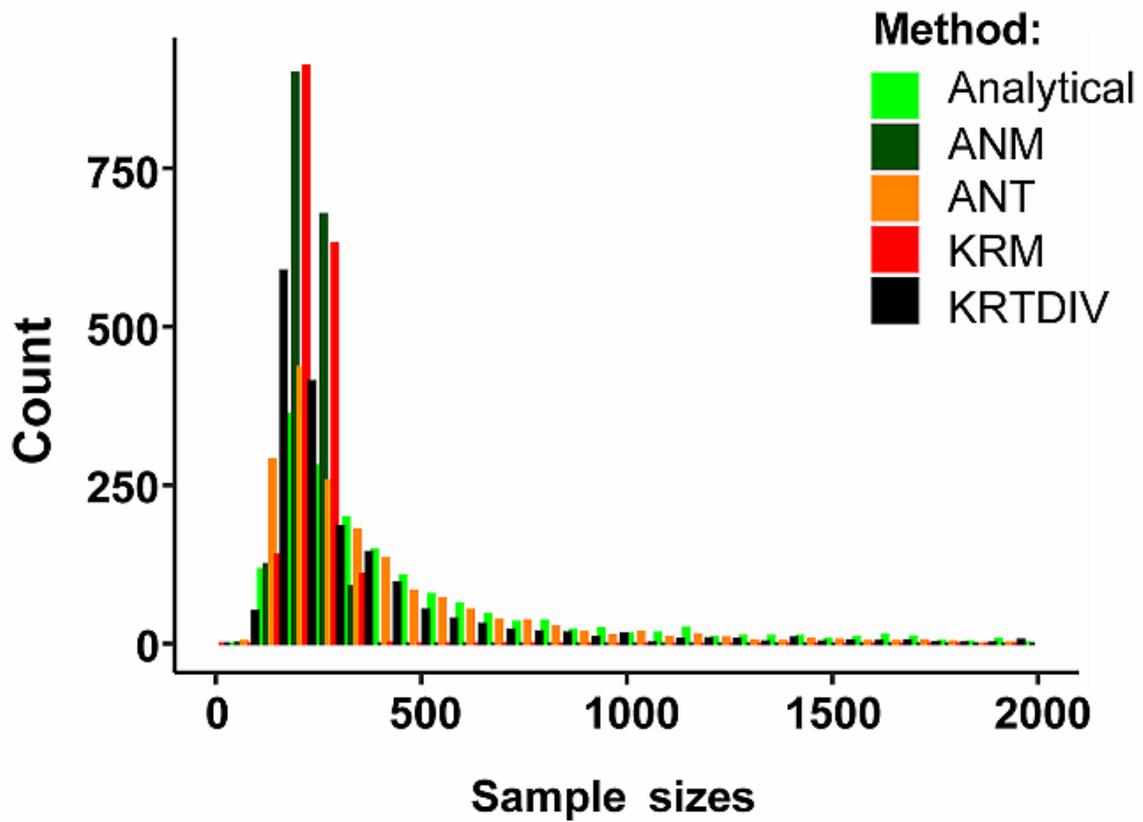

**Figure OR1.** Counts of predicted and measured sample sizes for 80% power in simulation run_three. Power prediction: KRTDIV: Monte-Carlo Kruskal-Wallis power; ANT: Monte-Carlo ANOVA; Analytical: analytical ANOVA. Measured power: KRM: Kruskal-Wallis; ANM: ANOVA. For settings, see Fig 1.

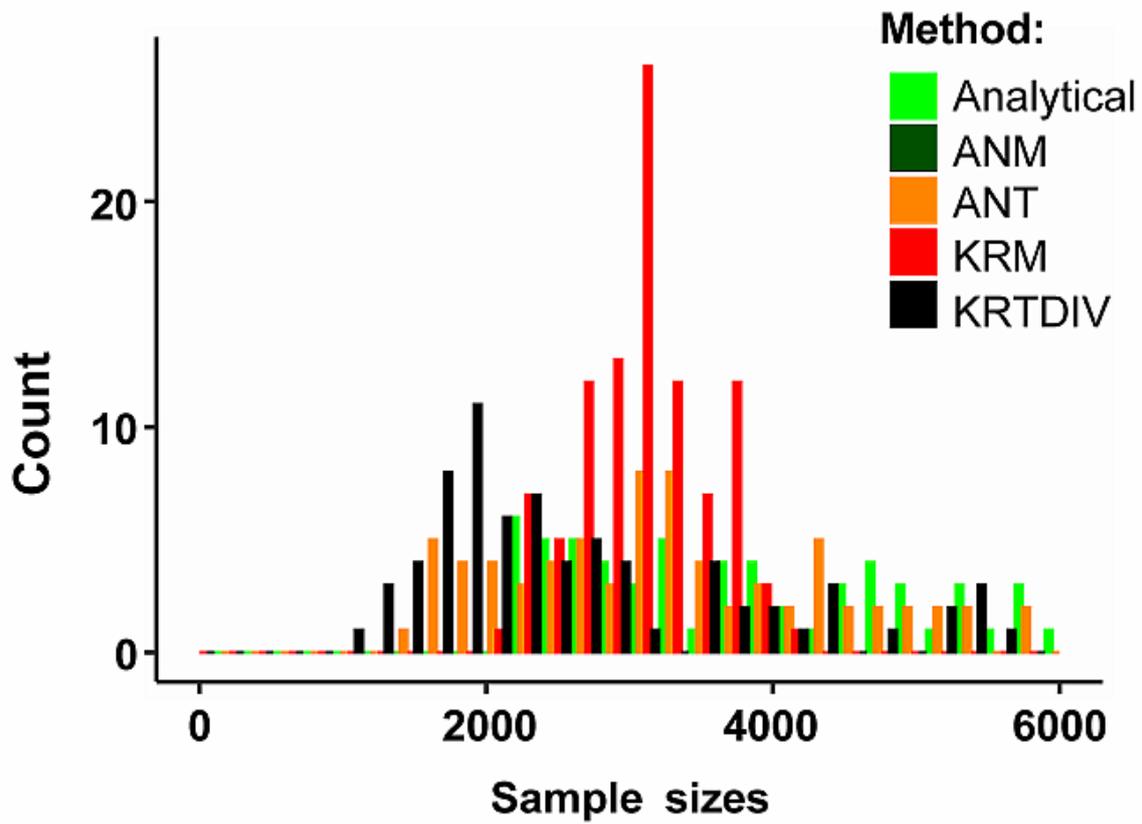

**Figure OR2.** Counts of predicted and measured sample sizes for power in run_DIAB: Diabetes/sleep hours study. For abbreviations see Fig OR1.

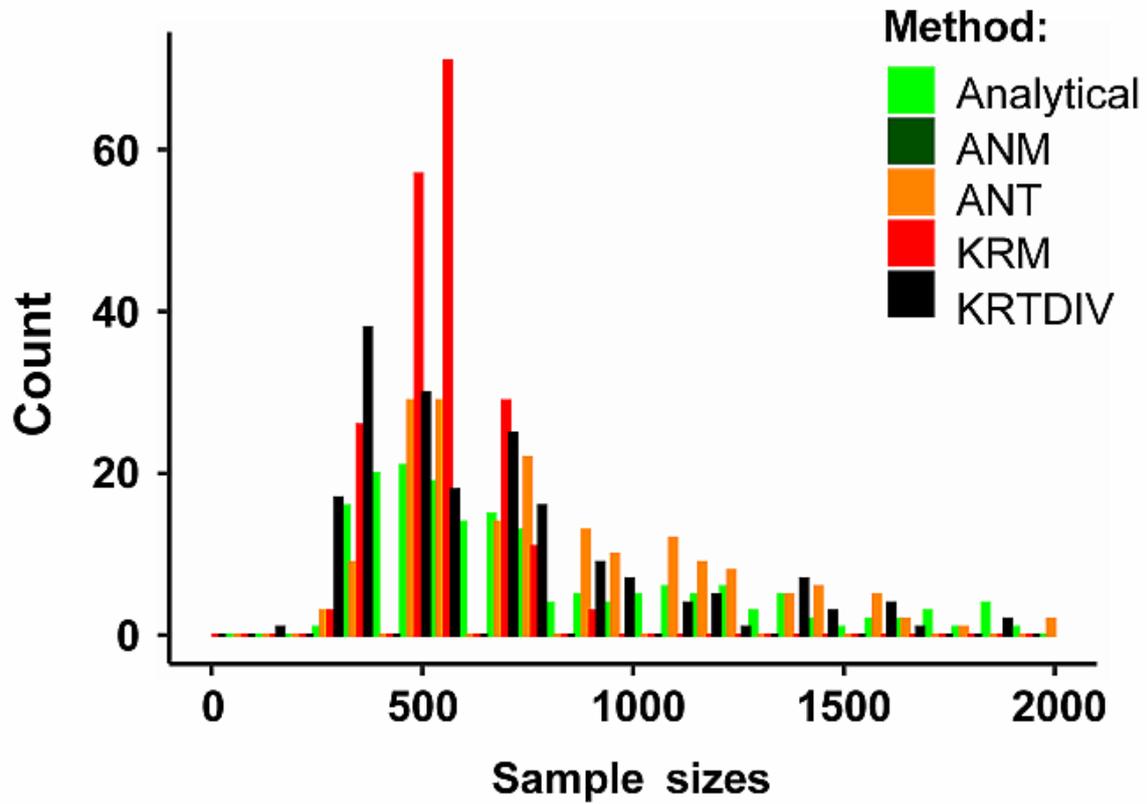

**Figure OR3.** Counts of predicted and measured sample sizes for power in run_Dial: Dialysis/systolic blood pressure study. For abbreviations see Fig OR1.

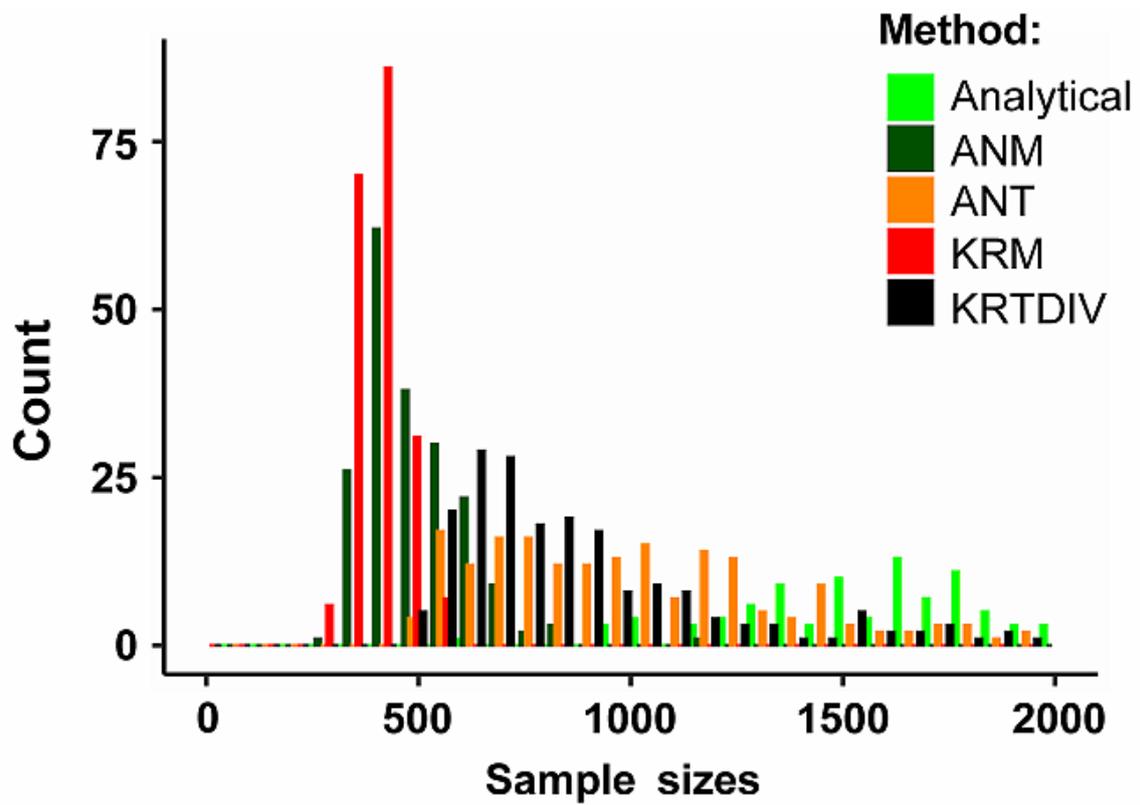

**Figure OR4.** Counts of predicted and measured sample sizes for power in run_Marrelaxed: Marital status/high-density-lipoprotein cholesterol study with relaxed ANOVA conditions. For abbreviations see Fig OR1.

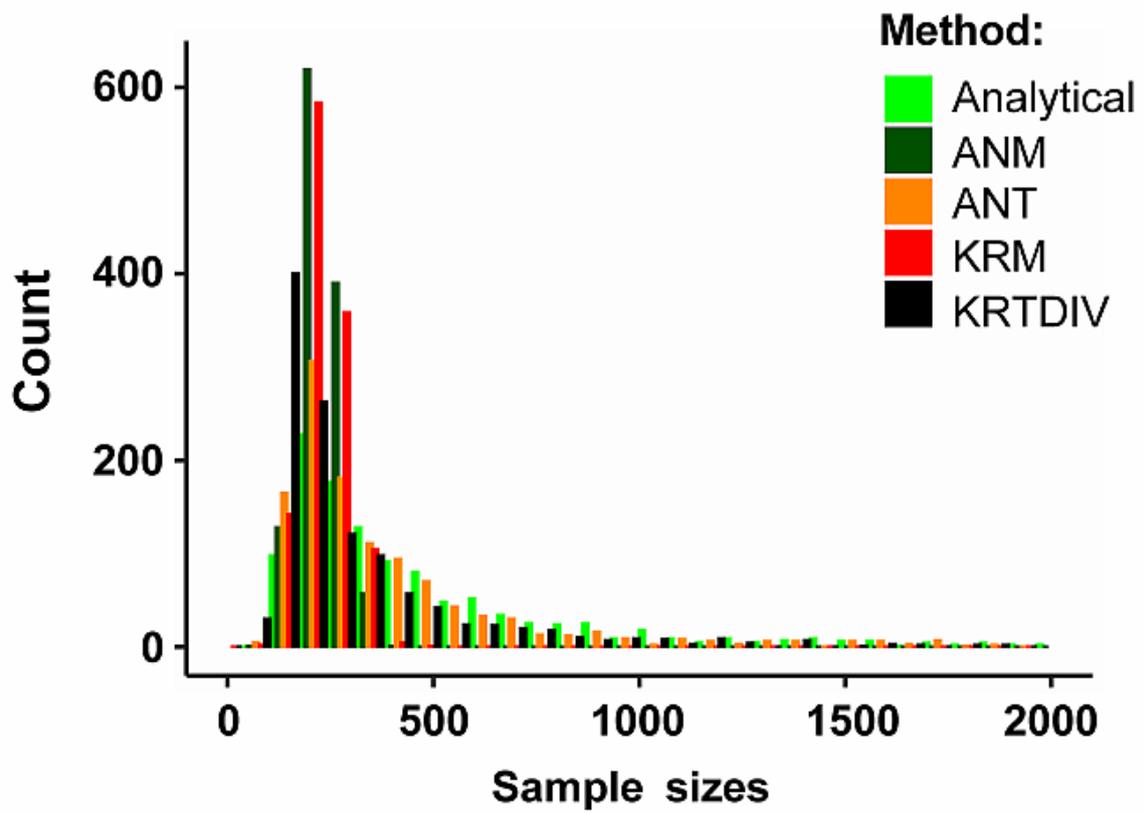

**Figure OR5.** Counts of predicted and measured sample sizes for power in simulation run_one. For abbreviations see Fig OR1; for settings see Fig 2.

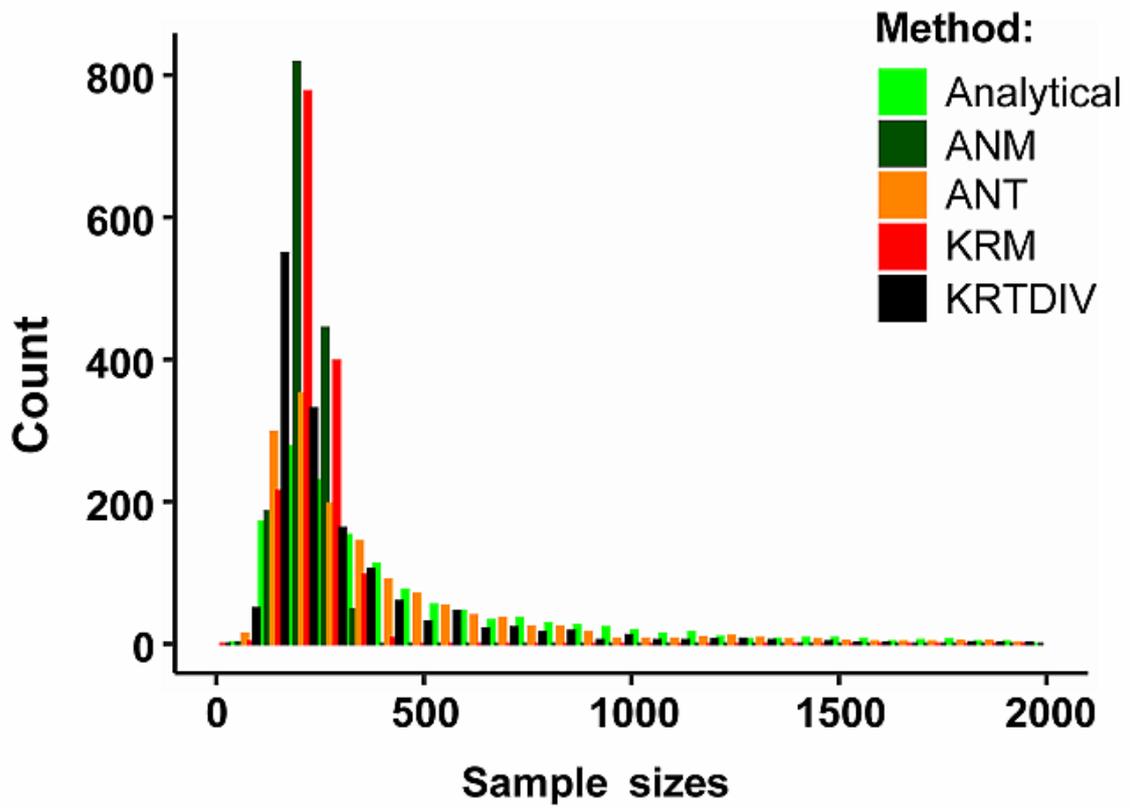

**Figure OR6.** Counts of predicted and measured sample sizes for power in simulation run_one with relaxed ANOVA conditions. For abbreviations see Fig OR; for settings see Fig 3.

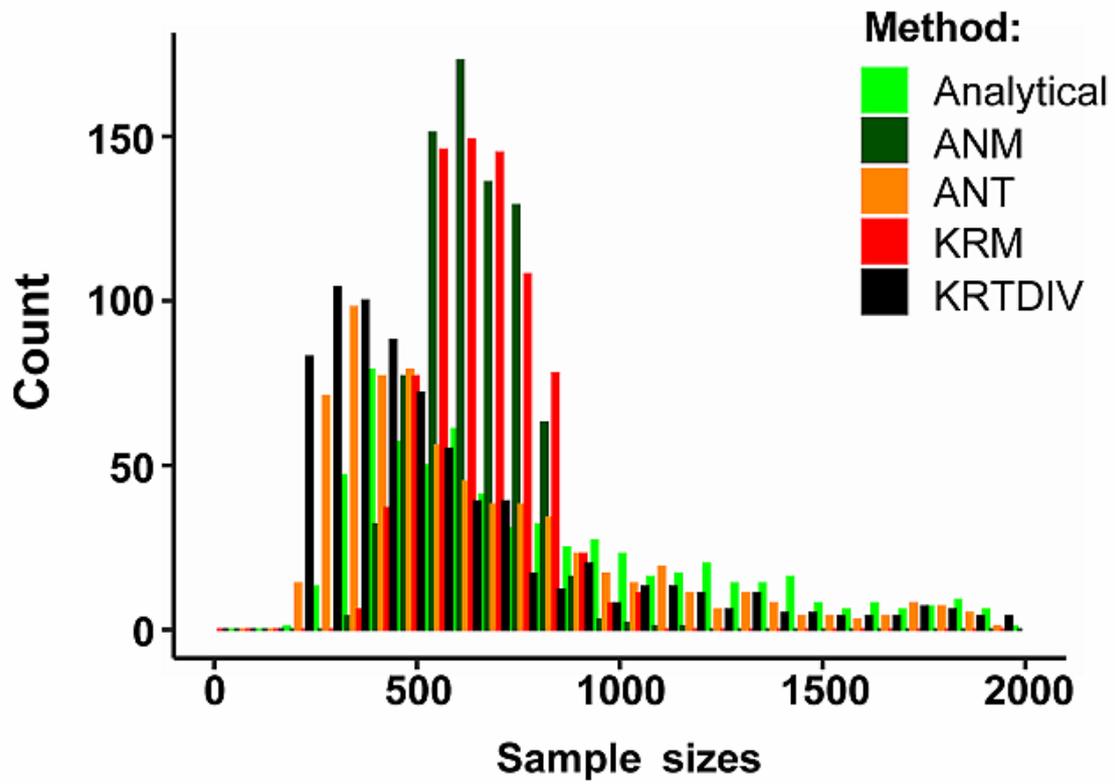

**Figure OR7.** Counts of predicted and measured sample sizes for power in simulation run_two. For abbreviations see Fig OR1; for settings see Fig 4.